# On a Japanese Subjective Well-Being Indicator Based on Twitter data.

Tiziana Carpi*     Airo Hino†     Stefano Maria Iacus‡§     Giuseppe Porro¶

**Abstract**

This study presents for the first time the SWB-J index, a subjective well-being indicator for Japan based on Twitter data. The index is composed by eight dimensions of subjective well-being and is estimated relying on Twitter data by using human supervised sentiment analysis. The index is then compared with the analogous SWB-I index for Italy, in order to verify possible analogies and cultural differences. Further, through structural equation models, a causal assumption is tested to see whether the economic and health conditions of the country influence the well-being latent variable and how this latent dimension affects the SWB-J and SWB-I indicators. It turns out that, as expected, the economic and health welfare is only one aspect of the multidimensional well-being that is captured by the Twitter-based indicator.

## 1 Introduction

The improvement of social well-being explicitly entered the policymaker agenda a few decades ago, when it became clear that objective measures of observable quantities - above all, the GDP - were unsatisfactory proxies of the welfare conditions of a community (Stiglitz et al., 2009). As a consequence, instruments for measuring and monitoring social well-being began to appear in the toolbox of policymakers, progressively moving the focus from objective to subjective evaluation: multidimensional indicators encompassing both objective and subjective dimensions of well-being (Barrington-Leigh and Escande, 2018; Fleurbaey, 2009), face-to-face or telephone surveys investigating samples of citizens about their own perception of quality life (Kahneman et al., 2004; Schwarz and Strack, 1999) and, after the development of the Internet, the application of of several techniques to the analysis of individual and collective mood through large-scale data provided by social networking sites (SNS), with the aim of drawing an evaluation of well-being status from conversations (Luhmann, 2017; Scollon, 2018) or word search (van der Wielen and Barrios, 2020) on the web.

In this context, Twitter is one of the most popular SNS, with 330 million monthly active users worldwide in 2019[1]. Due to the brevity of the messages allowed and to the huge amount of tweets potentially available and continually updated, the platform has been considered one of the most suitable information sources to estimate the emotional well-being, i.e. the "mood" or short-run component of the life quality evaluation.

Recent literature provides some examples of well-being evaluations that rely on Twitter data and, based on sentiment analysis methods, aim at monitoring the day-by-day evolution of self-declared emotional status of a community.

In particular, Dodds et al. (2011) built an happiness indicator, called *hedonometer*, based on a so-called 'closed vocabulary' approach: they measured the frequency of use of a set of ten thousand words for which they have obtained happiness evaluations on a nine-point scale, using Mechanical Turk[2]. Their dataset was huge, being made of around 4.6 billion expressions posted by over 63 million Twitter users from September 2008 to September 2011. The project is still ongoing and the hedonometer is now evaluated daily by the University of Vermont Complex Systems Center, which can provide, therefore, a time series since 2008[3].

A subjective well-being indicator - named *Gross National Happiness index* - has been proposed by Rossouw and Greyling (2020): the indicator is evaluated since 2019 in three Commonwealth member countries: South Africa, New Zealand and Australia[4]. The aim of the project is to measure, in real time, the sentiment of countries' citizens during different economic, social and political events: its first application has been an examination of the well-being impact of social restrictions imposed during the first wave of Covid-19 pandemic in South Africa (Greyling et al., 2020). In order to calculate the index, the sentiment analysis is applied to a live Twitter-feed, and each tweet is assigned either a positive, neutral or negative sentiment. Then, an algorithm evaluates an happiness score on a 0-to-10 scale. The Gross National Happiness index provides an happiness score per hour for each of the three countries.

---

*Department of Studies in Language Mediation and Intercultural Communication University of Milan, Milan, Italy.
†School of Political Science and Economics, Waseda University, Tokyo, Japan
‡Corresponding author, Email: stefano.iacus@ec.europa.eu
§European Commission, Joint Research Centre, Via Enrico Fermi 2749, 21027 Ispra (VA), Italy
¶Department of Law, Economics and Culture, University of Insubria, Como, Italy.

[1] https://statista.com
[2] https://www.mturk.com/
[3] https://hedonometer.org/timeseries/en_all/
[4] https://gnh.today

Several other studies apply sentiment analysis to the data provided by Twitter, in order to monitor short-run levels of happiness (Bollen et al., 2017), but also life satisfaction, defined as a medium-long run evaluation of life quality (Schwartz et al., 2013; Yang and Srinivasan, 2016; Lim et al., 2018; Durahim and Coşkun, 2015; Abdullah et al., 2015; Quercia et al., 2012; Greco and Polli, 2020).

An algorithm for sentiment analysis, named *Integrated Sentiment Analysis* (iSA) (Ceron et al., 2016) is used in this work to obtain a composite subjective well-being indicator for Japan, named SWB-J (*Subjective Well-Being Japan*). The advantage of iSA, compared to the wide range of sentiment analysis algorithms and methods applied to SNS big data repositories, is in that iSA is a human supervised machine learning method, where a sample of texts (training set) is first read and manually classified by human coders, and then the rest of the corpus (test set) is automatically classified by the algorithm. This allows for extracting qualitative information from a text without relying on dictionaries or special semantic rules: on the contrary, iSA can investigate cultural, psychological and emotional aspect of language, grasping all the nuances of informal and colloquial expressions. The feature is particularly significant because the analogous of SWB-J has been estimated for Italy (SWB-I) (Iacus et al., 2019, 2020a,b) with the same methodology: a comparison between the two indicators may be attempted, contributing to the challenging task of disentangling differences in life quality evaluations from cultural and linguistic specificities in expressing and communicating feelings and mood.

The SWB-J Project was created in collaboration between the University of Milan and Insubria in Italy and the two Japanese counterparts: The University of Tokyo and the Waseda University.

The paper is structured as follows: Section 2 is an introduction to the big data approach in the study of well-being and Section 3 briefly presents some cultural aspects of how emotions are expressed in Japanese language and reviews some recent literature on computational linguistic about extracting emotions from Japanese tweets. Section 4 describes the SNS data used in this study as well as other sources of data used in the subsequent sections while Section 5 describes the sentiment analysis methodology used to create the SWB-J indicator. Section 6 discusses the resulting new SWB-J indicator and compares its features with the Italian counterpart SWB-I. Section 7 presents a cross-country analysis aiming at explaining what can potentially impact the different patterns of SBW-J and SWB-I trough an econometric analysis. Section 8 summarizes the results and limits of this approach, and the Appendix contains additional technical material about the construction of the index.

## 2 Why to Estimate Subjective Well-Being via SNS' Big Data

Well-being evaluation has turned from the estimation of objective quantities to the estimation to the assessment of subjective mood and state of mind because observable variables - even in a multidimensional approach - have proven inadequate to accurately account for the welfare conditions of a society (Kuznets, 1934; Sen, 1980). Criticism to this approach has come to question its empirical relevance and opened doors to an overturning in the strategies for well-being evaluation: if both one-dimensional and multidimensional measures are unreliable, due to the limits of observable variables, the only feasible option to estimate individual and collective well-being is to explicitly *ask people* to express an evaluation about their own condition.

To this aim, surveys and questionnaires have been increasingly and widely used to collect information about well-being levels and dynamics of individuals and communities. Different methods to conduct surveys have been developed - also conditioned by the technology applied (face-to-face interview, telephone, internet) - in order to disentangle the incidental, emotional aspect of self-reported well-being and the evaluation of life satisfaction, which requires to examine current and past events in a medium or long-run perspective.

Though, survey-based researches have a significant drawback, that is the bias induced in well-being evaluation by the survey itself. It is a sort of "Hawthorne effect"[5] that Angus Deaton (Deaton, 2012; Deaton and Stone, 2016) pointed out: in fact, changing the order of the questions of a survey may be sufficient to affect the evaluation the respondents give about their own mood or quality of life. More generally, when the respondents are aware of being asked for an assessment of their own life and of being observed while giving the evaluation, the answer they give may be biased by this awareness. Therefore, the dilemma the analysts face is quite clear: on one hand, they wish *to ask* people for a self-evaluation of their well-being, in order to overcome deficiencies due to measurements based only on observable quantities; on the other hand, they should *not to ask* people for a self-reported evaluation, in order to avoid biases due to the awareness of respondents.

With the beginning of the era of virtual communication, a new source of large-scale dataset is provided that seems to address the need for this kind of information: in fact, the availability of a huge and continually updated flow of conversations on SNS theoretically provides a real-time opportunity to know what people think about the quality of their own daily life - both from an emotional and an evaluative perspective - without submitting any explicit questionnaire. This is fostering a stream of studies whose aim is to extract meaningful information from the enormous amount of words or images posted on well-known platforms such as Facebook, Twitter, Instagram (Voukelatou et al., ming).

---
[5]Also known as "observer effect", it is the phenomenon by which individuals modify their behavior in response to their awareness of being observed (Landsberger, 1958).

One of the main advantages of large-scale datasets coming from SNS is their continuous updating. It offers the opportunity for *nowcasting*[6] activity: in fact, while variables that are more traditionally assumed to be related to welfare - such as GDP or morbidity rates - are observable only with a time lag, that sometimes makes the policymaker intervention less effective, SNS data allow for a real-time monitoring of public sentiment and can anticipate changes in objective variables. Moreover, when the methods for sentiment analysis are language-independent (i.e. they can be applied to texts expressed in different languages, without any particular limitation), a comparison among linguistic and socio-cultural contexts is made possible, where not only differences in the use of language can emerge, but also cultural specificities - such as social conventions that impose a more strict self-control in expressing emotions - can be discovered, as far as they are recorded in virtual conversations.

On the other hand, SNS data have also some intrinsic *limitations*. First of all, users of these platforms are not a representative sample of the whole population: therefore, any social well-being evaluation achievable from the analysis of these data cannot be immediately extended to the whole population. Adjusting procedures can be applied to make the results more general but, above all and despite their limited representativeness, SNS can be considered a sort of opinion-making arena, where expressed ideas affect or anticipate collective sentiment and trends. This, actually, suggests that a second drawback can be imputed to evaluation of well-being via SNS data: the use of social networks itself can alter self-perceived or self-declared well-being. In fact, even if SNS users do not answer any explicit question about their own personal status, they are aware they are sharing their feelings with a community and this can distort their well-being self-report, in order to satisfy self-representation needs. Furthermore, SNS messages and texts seem more suitable to reflect short-term mood changes than a long-term evaluation of life quality: therefore, a well-being indicator based on SNS data would be more reliable as a measure of emotional well-being than a source of life evaluation. However, despite the validity of this remark, adequate statistical analysis can help in separating the volatile and the structural components of well-being path described by virtual conversations on the net.

An important issue raised by the availability of this new data source is a technological one: the increase in computational power of technological devices does not guarantee, *per se*, the ability to separate helpful information from background noise in virtual conversations. Following Gary King, we can say that "Big data is not about the data"(King, 2016): it is rather about the opportunity to extract knowledge from them, and this requires adequate methodologies and tools. Fortunately, recent advancements in statistical theory and its applications are improving the capacity of social scientists to analyze the content of these large-scale dataset and promoting the dissemination of different methods of sentiment analysis. The subjective well-being indicator we propose in this work is based on the iSA algorithm, which is one of these new methods.

## 3 Expressing Emotions in the Japanese Culture and SNS

As mentioned by Miyake (2007), in traditional Japanese communication, people tend to maintain distance and make sure and subjective experiences (Matsumoto, 1999). On textual analysis, most studies focus on the identification of big corpora of web blogs or SNS posts (Ptaszynski et al., 2014) in the context of sentiment and affect analysis. As emoticons, or emoji, are peculiar in Japanese written digital communication. For example, before the graphical emoticons appears, while in the western cultures horizontal emotions like ":)" were in use, in Japan (an other Asian countries) emoticons were and are traditionally vertical, like "(`o´)". Emoji's were already installed as a standard package of messaging platforms of mobile devices in Japan in the late 1990's (e.g. Jphone and iMode). The development of emoji is distinctive in Japan and arguably originates from the 'kanji' culture where characters represent an idea or concept as a graphic symbol (which also applies to Asian countries that employ Chinese characters). Despite the abundance of emoticons, a large cross-country study seems to prove that regardless of the culture, vertical and horizontal emoticons convey similar concepts (Park et al., 2014). Still, emoticons coupled with adverbs seems to be able to predict better than simple emoticons the affective perception of a text message (Rzepka et al., 2016). Many other studies related to the association between emoticons and emotions can be found in the literature (e.g. Shoeb and de Melo, 2020; Novak et al., 2015) but they are not specific to the Japanese language. In relation to the Japanese culture to express emotions by non-verbal means, a graphic design known as the ASCII art has been quite extensively used in a bulletin board such as 2channel (one of the most popular online bulletin boards). Personality trait estimation of Japanese Twitter accounts have been studied in Kamijo et al. (2016). Large scale studies on automatic sentiment tagging in Japanese during crisis periods can be found in (Vo and Collier, 2013). More linguistic analysis on specific Japanese terms related to likeness and happiness have recently appeared (for the word kawaii see, e.g. Iio, 2020) as well as gender-specific language studies (Carpi and Iacus, 2020).

In summary, all the current studies are either dictionary or semantic rule based or they apply some version of the classical Word2Vec approach (Mikolov et al., 2013) or standard NLP techniques (e.g. Bengio et al., 2003). In this study, as it will be described in Section 5, we use a mixed qualitative and quantitative approach which tries to take into account the complexity of the well-being dimensions and the language used to express it. Indeed, the analysis does not focus on special features of a message but on the whole set of the words in a tweet after accurate training by humans which are Japanese mother language natives and fully understand all the shadings of the natural language used to express emotions.

---

[6]The term is a contraction for "now" and "forecasting": it refers to the opportunity to collect information about some quantities or variables in real time.

# 4 Data Collection

The data used in this work come from two different repositories that were collected under two different projects but in both cases using Twitter search API. The Japanese tweets were collected using only the filter on language = `Japanese` and country = `Japan` and similarly for Italy (`Italian` and `Italy`). It is worth to mention that Twitter posts do not belong to individuals randomly chosen from a physical population (Baker et al., 2013; Murphy et al., 2014). The reference population is the population of posts of all Twitter accounts selected in the analysis. Moreover, Twitter accounts cannot be uniquely associated to individuals and some accounts are more active than others. For these reasons, the focus of our analysis is on the total volume of the posts collected (in Japan, written in Japanese language, during the reference period) through the public Twitter "search" and "streaming" API[7]. As per the official documentation, Twitter search API only provides a 10% sample of all tweets though the company does not disclose any information about the representativeness of the sample with respect to the whole universe of tweets posted on the social network. Nevertheless, according to our personal experience, also confirmed in large scale experiments by Hino and Fahey (2019), the coverage of topics and keywords is quite accurate and appears to be randomly selected: therefore we consider the Twitter data used in this study as a representative sample of what is discussed on Twitter. According to Statista[8] there are about 8 million accounts active daily in Italy whilst about 52 millions are in Japan, therefore the number of tweets posted is not comparable. Despite these limitations, the advantage of using Twitter data is that the collection of data can be done in (almost) continuous time and, moreover, instead of asking something through a web-form, thanks to the human supervised qualitative analysis explained in Section 5 it is possible to capture expressions of well-being from the texts directly.

For Italy, the original project collected about 250.4 millions of tweets in the period 01-02-2012/21-06-2018, with a median of 50,000 tweets per day. For Japan, for several technical reasons, we were able to collect at most 50,000 tweets a day, amounting to about 60.8 millions of tweets in the period 24-08-2015/31-12-2018. Table 2 reports summary statistics. In the same table, we report also the values of the Happy Planet Index (HPI) by New Economics Foundation (2016) and the Human Development Index (HDI) by the United Nations Development Programme (2019), for the same years available for the SWB-I an SWB-J indexes. Data are taken from the data provider TheGlobalEconomy[9] as the other economic variables listed in Table 1, mostly coming from The World Bank, the International Monetary Fund, the United Nations, and the World Economic Forum. We also collected from OECD[10] the variable Life Expectancy of Males at 40, to capture the perception and quality of aging. This is the only yearly variable we consider as it varies non-monotonically in time and, moreover, is also correlated positively with economy growth in Italy ($\rho = 0.91$) but negatively in Japan ($\rho = -0.27$).

| Variable | Frequency | description |
|---|---|---|
| GDP growth | quarterly | Percent change in quarterly real GDP year on year |
| Consumption growth | quarterly | Percent change year on year |
| Investment growth GDP | quarterly | Percent change year on year |
| Unemployment rate | monthly | percentage of work force |
| Life Expectancy at 40 | yearly | male only |
| Happy Planet Index | yearly | |
| Human Development Index | yearly | |

**Table 1:** Economic and environmental variables used in the econometric analysis of Section 7 plus two additional well-being indexes.

# 5 How to Extract Subjective Well-Being from Tweets

The SWB-J index is a multidimensional well-being indicator whose components were inspired by the dimensions adopted by the New Economic Foundation think-tank for its Happy Planet Index (New Economics Foundation, 2016). In summary, the SWB-J mimic the same indicator SWB-I previously built for Italy (Iacus et al., 2019, 2020b,a) and consists of eight dimensions that concern three different well-being areas: personal well-being, social well-being and well-being at work. More in detail:

1. *Personal well-being*:
    - **emotional well-being**: the overall balance between the frequency of experiencing positive and negative emotions, with higher scores showing that positive feelings are felt more often than negative ones (`emo`);

---

[7]Application Programming Interface

[8]https://statista.com

[9]TheGlobalEconomy.com

[10]https://data.oecd.org/healthstat/

- **satisfying life**: having a positive assessment of one's life overall (`sat`);
- **vitality**: having energy, feeling well-rested and healthy while also being physically active (`vit`);
- **resilience and self-esteem**: a measure of individual psychological resources, of optimism and of the ability to deal with life stress (`res`);
- **positive functioning**: feeling free to choose and having the opportunity to do it; being able to make use of personal skills while feeling absorbed and gratified in daily activities (`fun`);

2. *Social well-being*:
   - **trust and belonging**: trusting other people, feeling treated fairly and respectfully while experiencing sentiments of belonging (`tru`);
   - **relationships**: the degree and quality of interactions in close relationships with family, friends and others who provide support (`rel`);

3. *Well-being at work*:
   - **quality of job**: feeling satisfied with a job, experiencing satisfaction with work-life balance, evaluating the emotional experiences of work and work conditions (`wor`).

It has been known that personal well-being is closely associated with social well-being and well-being at work in Japan compared to the United States (Kitayama et al., 2000; Ford et al., 2015), and most recently China (Wong et al., 2020). Given the collectivist nature of the Japanese society, it makes sense to measure subjective well-being based on multiple dimensions encompassing social well-being and well-being at work. Asian countries tend to mark a lower score in reported subjective well-being (Diener et al., 1995) but few work has attempted to measure the level of subjective well-being with social media data. With the multi-dimensional nature of the index, one can better understand and elucidate an internal mechanism of subjective well-being. To extract semantic meaning from tweets, in this study we use supervised sentiment analysis and, in particular, the iSA (Integrated Sentiment Analysis) algorithm (Ceron et al., 2016) which has been also used to capture various aspects of happiness and well-being from Twitter data (Curini et al., 2015). iSA is a human supervised machine learning method, where a sample of texts (training set) is first read and manually classified by human coders, and then the rest of the corpus (test set) is automatically classified by the algorithm. The supervised part is essential in that this is the step where qualitative information can be extracted from a text without relying on dictionaries or special semantic rules but rather on cultural, psychological and emotional interpretation. Other approaches based on user-defined dictionaries exist, but mainly focusing on the concept of happiness (Bollen et al., 2011; Zhao et al., 2019). The advantage of iSA over other machine learning techniques, is that it is designed to estimate directly the aggregated distribution of the opinions (e.g., positive, negative, neutral) without passing through the individual classification of posts in the test sets. This approach vastly reduces the estimation error. Moreover, being iSA a sequential method, in this context of highly noised data, the size of the training set needed to reach the same accuracy of other methods is usually smaller by a factor of 10 or 20 times. The reader can refer to Ceron et al. (2016) for the technical explanation of the method.

Note that, in discussing the concept of well-being in the Japanese society, Kumano (2018) distinguishes the two types of well-being: the *shiawase* or hedonic well-being, and *ikigai* or eudaimonic well-being. Clearly, both are captured in our analysis but not distinguished as this will be the scope of future work.

The data for the training set are extracted according to the filtering keywords are listed in Tables 10, 11, 12, 13, 14, 15 and 16 of the Appendix. Remark that, even though the training set is built by filtering the data, the whole statistical analysis is done on the complete repository of tweets collected.

During the qualitative step of the training set classification, coding rules have been distributed to human coders:

- The first general rule is mark/tag/code `Off-Topic` posts appropriately. At this stage the machine learning algorithm will understand *noise*.
- The second rule is: if you are not fully convinced about the semantic context of a post do not classify it, just skip it go to the next one. These are not `Off-Topic`, let the algorithm try to classify it for you.
- It is admissible to classify `RT` (re-tweet) as original tweets from the account. This is an assumption of *transfer of emotion/opinion* which we assume to be the same as if they were expressed by the user account directly. Other researcher may disagree with this assumption of course.
- As each tweet can be classified along one or more dimensions, always try to consider parallel coding for all the categories and leave unanswered/untagged those who do not apply to a given category.

Examples of coding rules and real tweets classifications are given in the Appendix.

Once the training set has been completely hand-coded, the iSA algorithm is applied to daily test sets of data. Each estimated distribution will contain the entries `positive`, `neutral`, `negative` and `OffTopic`. The `OffTopic` category represent the daily *noise* in the data, the rest represent the *signal*. For each component of the index, for example `emo`, the index is calculated, for day *d*, as follows:

$$\text{emo}_d := \frac{\%\text{positive}}{\%\text{positive} + \%\text{negative}} \in [0,1] \tag{1}$$

The rationale behind formula (1) is that the intent is to capture expressions or judgments, and this is why comments not expressing any issue, judgments, etc, i.e. those classified as `neutral`, are removed from the calculation. The day $d$ is easily extracted from the timestamp of Twitter data which is always present in the metadata of a tweet.

Finally, the SWB-J index is the simple average of the eight components `emo`, `sat`, `vit`, `res`, `fun`,`tru rel` and `wor`. Although it falls out of the scope of this study, specific weighted averages of the components might be considered in order to obtain a well-being indicator oriented to specific life dimensions.

# 6 Preliminary Analysis of the SWB-I & SWB-J Indexes

Table 2 shows the yearly average values of SWB-I and SWB-J since 2012 for Italy, and since 2015 for Japan. What we notice is that the Japanese indicator shows a high medium-run stability in the range $[52.5, 54.5]$. In the Italian case, on the contrary, the medium-run variability is higher - the range of yearly values is $[48.7, 57.7]$ and the value of SWB-I is, in some years, significantly lower or significantly higher than SWB-J. On the other hand, the standard deviations of the two indicators suggest (and the inspection of Figure 1 proves) that the short-run volatility of SWB-J is definitely higher, compared to SWB-I. These results depict Japanese Twitter users as more reactive to the day-by-day events and emotions, while their evaluation of quality life is, on average, stable around quite satisfactory values. To be more exact, the Italian indicator show a a high variability period, which basically coincides with the year when Italy organized and hosted the Expo 2015 event: that was an abnormal time frame, when heavily negative (due to controversies over delays in the preparation of the event and to allegation of bribery to some of the organizers) and strongly positive feelings (due to the appreciation and success of the event) rapidly emerged and changed in the Italian public opinion[11].

An examination of the single components of SWB-J in Table 4 confirms the stability - along with a slight decline - of Japanese subjective well-being. A first paradox emerges looking at the well-being perception about social relationships: the outstanding value of the sub-component evaluating the quality of family relationship and friendship (`rel`) is, to some extent, at odds with the perceived well-being in terms of trust and sentiment of belonging (`tru`): this may indicate that the positive feelings nourished towards family and friends are not generalized to the rest of the society. The emotional sub-component of SWB-J is consistent with the global indicator, as the other dimensions related to personal well-being: slightly above the average we find the `fun` component, regarding the opportunity to do and to choose, and the involvement and satisfaction in daily activity; slightly below the self-perception of health and physical vitality (`vit`). Also the subjective well-being at work (`wor`) is strictly in line with the average SWB-J: this is in contrast with the Italian case (see Table 3), where the `wor` component is the most volatile and does not show strong correlation with the overall value of SWB-I, and likely documents the strong identification Japanese people feel between their satisfaction as workers and their global well-being or that concerns at work are not expressed for cultural reasons which is not unexpected.

A look at the correlation between SWB-J (and SWB-I) and two well-known well-being indicators may raise some concerns. Table 5 shows a high correlation of SWB-J with the Happy Planet Index (HPI), developed for the first time in 2006 by the New Economics Foundation (2016). The HPI aims at giving a measure of sustainable well-being: it compares how efficiently residents of different countries are using natural resources to achieve long, high well-being lives. On the other hand, SWB-J is negatively related (with a correlation index equal to $-0.99$) to the Human Development Index, elaborated since 1990 by the United Nations Development Programme (UNDP), according to Amartya Sen's capability approach to well-being definition and evaluation (Robeyns, 2006). In measuring well-being, HDI takes into account three dimensions: health, education and material standards of living. It can be noted that the Italian SWB-I is positively related to both the indicators: a weak relation is shown with HPI and a strong one with HDI. All this should remind us that the plethora of well-being indices currently available seldom gives a measure of the same variable: each indicator addresses a specific definition of well-being and the relationships among all these definition are sometimes unclear and ambiguous. This does not imply that the measures provided are wrong or unreliable: it only require extreme clearness in explaining the methodology followed to construct the indicator, the data source, the definition of well-being the indicator aspires to account for.

# 7 Cross-Country Analysis 2015-2018

In this section we focus the attention on the impact of different economic variables on the SWB indicators. We make use of monthly and quarterly data of Table 1 and interpolate quarterly data at monthly frequency to make use of as much data as possible. Note that, while Italy is examined over the period 2012-2015, the analysis is restricted to the period 2015-2018 when both countries are considered together for comparison purposes. For the analysis we used the Structural Equation Modeling (SEM) with continuous response variable (Bollen, 1989) approach. SEM is a common method to test complex relationships between dependent variables, independent variables, mediators, and latent dimensions. We assume that the true well-being is a latent variable influenced by the economic status of the country, which itself is supposed to be a latent variable, and by the health status of a country measured through the expectancy of life, and that the Twitter SWB-I/SWB-J indexes are observable measures of some aspects of the well-being latent variable.

---

[11] In fact, despite several positive spikes, the average SWB-I in 2015 is the lowest of the examined period.

| Year | 2012 | 2013 | 2014 | 2015 | 2016 | 2017 | 2018 |
|---|---|---|---|---|---|---|---|
| SWB-I | 48.9 | 52.2 | 49.7 | 48.7 | 50.5 | 57.7 | 55.7 |
|  | (4.2) | (3.8) | (4.9) | (9.8) | (7.5) | (4.5) | (7.1) |
| tweets | 44.2M | 40.8M | 34.4M | 38.3M | 55.2M | 32.6M | 14.9M |
| HPI | – | 6.02 | – | 5.95 | 5.98 | 5.96 | 6.00 |
| HDI | 0.874 | 0.873 | 0.874 | 0.875 | 0.878 | 0.881 | 0.883 |
|  |  |  |  |  |  |  |  |
| SWB-J | – | – | – | 54.4 | 53.6 | 53.2 | 52.5 |
|  | – | – | – | (13.4) | (11.1) | (13.1) | (12.7) |
| tweets | – | – | – | 6.5M | 18.2M | 18.2M | 17.8M |
| HPI | – | – | – | 5.99 | 5.92 | 5.92 | 5.92 |
| HDI | – | – | – | 0.906 | 0.910 | 0.913 | 0.915 |

**Table 2:** Yearly average values of SWB-I and SWB-J, their standard deviation in parentheses, and number of tweets in million. Data are in the period 01-02-2012/21-06-2018 for Italy and 24-08-2015/31-12-2018 for Japan. For the Happy Planet Index (HPI) and Human Development Index (HDI) source World Bank.

| Year | SWB-I | emo | fun | rel | res | sat | tru | vit | wor |
|---|---|---|---|---|---|---|---|---|---|
| 2012 | 48.9 | 60.5 | 67.8 | 34.1 | 55.1 | 43.9 | 59.2 | 53.9 | 16.4 |
| 2013 | 52.2 | 57.3 | 73.3 | 37.4 | 57.2 | 55.0 | 64.0 | 58.0 | 15.5 |
| 2014 | 49.7 | 48.2 | 68.3 | 39.7 | 56.1 | 52.4 | 62.6 | 55.2 | 15.1 |
| 2015 | 48.7 | 53.1 | 52.7 | 57.7 | 55.4 | 33.2 | 37.7 | 57.0 | 42.8 |
| 2016 | 50.5 | 62.2 | 40.5 | 65.9 | 59.7 | 30.2 | 28.9 | 58.4 | 58.0 |
| 2017 | 57.7 | 23.5 | 59.1 | 64.4 | 45.8 | 79.0 | 20.2 | 80.6 | 88.9 |
| 2018 | 55.7 | 40.4 | 57.8 | 59.1 | 46.4 | 64.9 | 26.6 | 74.5 | 76.2 |

**Table 3:** The values of each components of the SWB-I index.

| Year | SWB-J | emo | fun | rel | res | sat | tru | vit | wor |
|---|---|---|---|---|---|---|---|---|---|
| 2015 | 54.4 | 54.8 | 59.3 | 75.2 | 54.4 | 56.9 | 35.4 | 43.2 | 55.9 |
| 2016 | 53.6 | 53.5 | 59.4 | 73.9 | 58.9 | 53.0 | 35.6 | 42.6 | 52.2 |
| 2017 | 53.2 | 51.0 | 57.9 | 75.7 | 55.9 | 51.5 | 36.1 | 42.7 | 55.0 |
| 2018 | 52.5 | 51.5 | 57.0 | 72.3 | 54.9 | 53.4 | 35.6 | 43.3 | 52.2 |

**Table 4:** The values of each components of the SWB-J index.

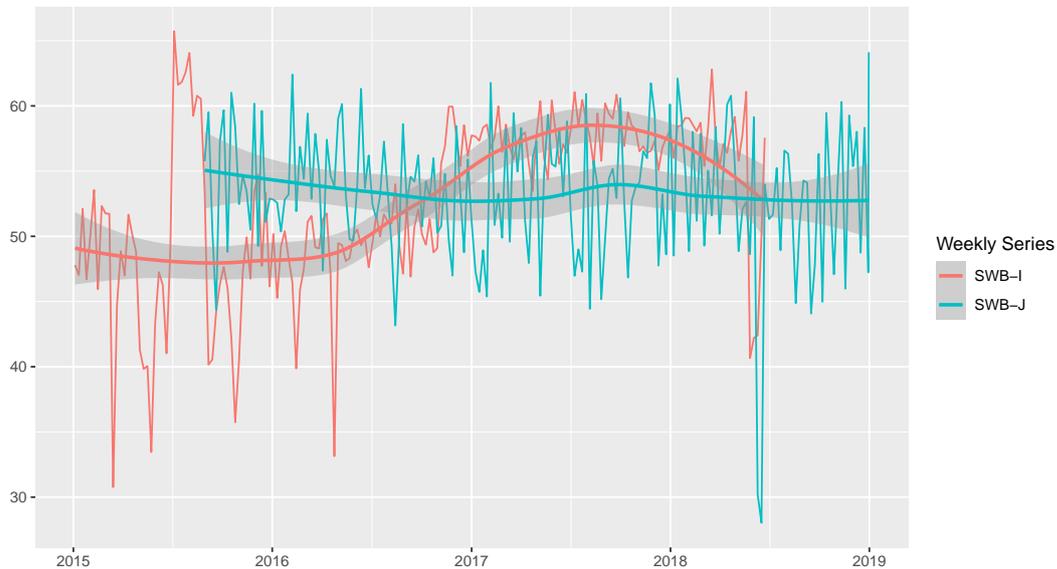

**Figure 1:** SWB-I and SWB-J weekly average series with estimated local-linear regression trends and standard errors bands. The peak in June-September 2015 for Italy is in correspondence of the Expo 2015 event.

|       | Happy Planet Index | Human Development Index |
|-------|--------------------|-------------------------|
| SWB-I | 0.14               | 0.80                    |
| SWB-J | 0.81               | -0.99                   |

**Table 5:** Correlation between the yearly average SWB-I and SWB-J and the two indexes Happy Planet Index and Human Development Index.

In statistical terms, SEM consists of regression analysis, factor analysis, and path analysis to explore interrelationships between variables. It is a confirmatory technique where an analyst tests a model to check consistency between the relationships put in place. The following latent dimensions are theorized:

- Economy: captured by GDP growth, consumption growth, investment growth and unemployment rate;
- Well-being: we assume it is affected by the Economy latent variable and by the life expectancy taken as a proxy of the health conditions, and in turn the Well-being variable determines SBW-I/SWB-J measures. Then, a path diagram is constructed to represent inter-dependencies of the independent variables (GDP growth, consumption growth, investment growth, unemployment rate, Life Expectancy at 40), the latent dimensions, and the dependent variable (SWB-I/SWB-J):

$$\text{Economy} \mapsto \text{GDP growth} + \text{Consumption growth} + \text{Investment growth} + \text{Unemployment rate}$$

$$\text{Well-being} \leftarrowtail \text{Economy} + \text{Life Expectancy at 40}$$

$$\text{SWB-J/SWB-J} \leftarrowtail \text{Well-being}$$

Further, the residual correlation among the observed variables is also captured in the model among the variable GDP growth and Life Expectancy at 40, Consumption growth, Investment growth and Unemployment rate. The results of the fitted model are presented in Table 6, while Figures 3-4 give a graphical representation of the same fitting. The models have been fitted using the `lavaan` package (Rosseel, 2012) and plots have been generated through the `semPlot` package (Epskamp, 2019).

## 7.1 Interpretation of the SEM model

In the Italian case (see Figure 4 and Table 6 bottom panel), all the observed economic variables have an expected and significant relationship with the Economy latent variable which, in turn, positively and significantly affect Well-being. A few anomalies, on the contrary, may be noted in the analysis of Japan. First of all, we notice from Figure 3 and Table 6 (top panel) that the relationship between the observable economic variables and the Economy latent variable, as well as the inter-dependencies among the observable economic variables, have the expected sign. Moreover, both the investment growth and the consumption growth rate show a significant relationship, whose coefficients are higher compared to Italy: being investment and consumption the main components of the aggregate demand, this likely explains why the relationship with GDP growth comes out to be statistically non-significant. But, on the other side, the unemployment rate is negatively and significantly related to the state of the Economy. Nevertheless, the Economy latent variable does not significantly affect Well-being: this is probably our main result and suggests that well-being perception among the Japanese is not determined by the objective, observable and mainly economic variables that have been traditionally used to measure the welfare of a country. It is worth reminding here that Diener et al. (1995) observed the tendency of Asian cultures, compared to continental European and Anglo-Saxon ones, to mark a lower score in reported subjective well-being, economic conditions being similar, documenting probably that economic wealth is not necessarily the most important component of perceived well-being.

Life expectancy at 40, which we adopt as a proxy of public health conditions, shows an opposite relationship in Japan and in Italy both with the Well-being latent variable and the GDP growth rate. As shown in Figure 2, life expectancy is increasing in both countries over the period 2015-2108. Similarly, the Italian GDP growth rate has a positive trend in these years, whilst the Japanese one is more fluctuating: this justifies the observed different sign of the relationship.

Lastly, the SWB indicator is positively and significantly related to Well-being in both countries. Nevertheless, the coefficient is higher in the Japan case, highlighting that SWB-J more closely resembles the well-being level, as depicted by the latent variable.

|  | Relationship |  | Coefficient | Std.Err. |
|---|---|---|---|---|
| Japan 2015-2018 | | | | |
| Well-being | $\mapsto$ | SWB-J | 0.940*** | 0.101 |
| Economy | $\mapsto$ | Economic growth | 0.406 | 0.497 |
| Economy | $\mapsto$ | Unemployment rate | −0.377** | 0.148 |
| Economy | $\mapsto$ | Consumption growth | 1.173*** | 0.159 |
| Economy | $\mapsto$ | Investment growth | 0.730*** | 0.155 |
| Well-being | $\leftarrowtail$ | Economy | 0.178 | 0.123 |
| Well-being | $\leftarrowtail$ | Life expectation at 40 | −0.362** | 0.159 |
| Economic growth | cov | Life expectation at 40 | −0.743*** | 0.174 |
| Economic growth | cov | Consumption growth | 0.404 | 0.525 |
| Economic growth | cov | Investment growth | 0.597* | 0.358 |
| Economic growth | cov | Unemployment rate | −0.440** | 0.195 |
| Italy 2015-2018 | | | | |
| Well-being | $\mapsto$ | SWB-I | 0.597*** | 0.113 |
| Economy | $\mapsto$ | Economic growth | 0.514*** | 0.190 |
| Economy | $\mapsto$ | Unemployment rate | −0.581*** | 0.178 |
| Economy | $\mapsto$ | Consumption growth | 0.597*** | 0.178 |
| Economy | $\mapsto$ | Investment growth | 0.398** | 0.179 |
| Well-being | $\leftarrowtail$ | Economy | 0.921** | 0.375 |
| Well-being | $\leftarrowtail$ | Life expectation at 40 | 0.834*** | 0.242 |
| Economic.growth | cov | Life expectation at 40 | 0.246** | 0.123 |
| Economic.growth | cov | Consumption growth | 0.121 | 0.137 |
| Economic.growth | cov | Investment growth | 0.230 | 0.134* |
| Economic.growth | cov | Unemployment rate | 0.004 | 0.121 |

*Note:* *p<0.1; **p<0.05; ***p<0.01

**Table 6:** Estimated coefficients for the SEM model applied to the Japanese (top) and Italian (bottom) data for the period 2015-2018.

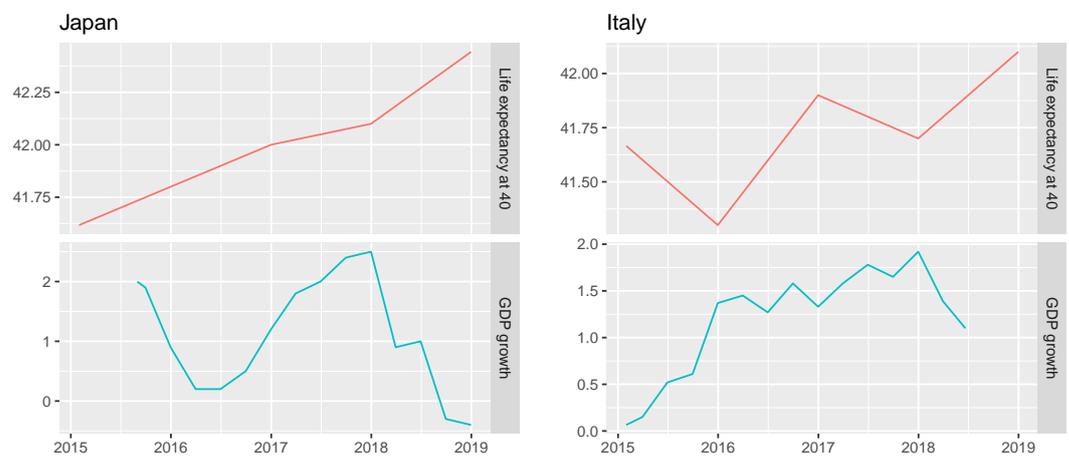

**Figure 2:** GDP growth and Life expectancy at 40 for Japan and Italy, period 2015-2018.

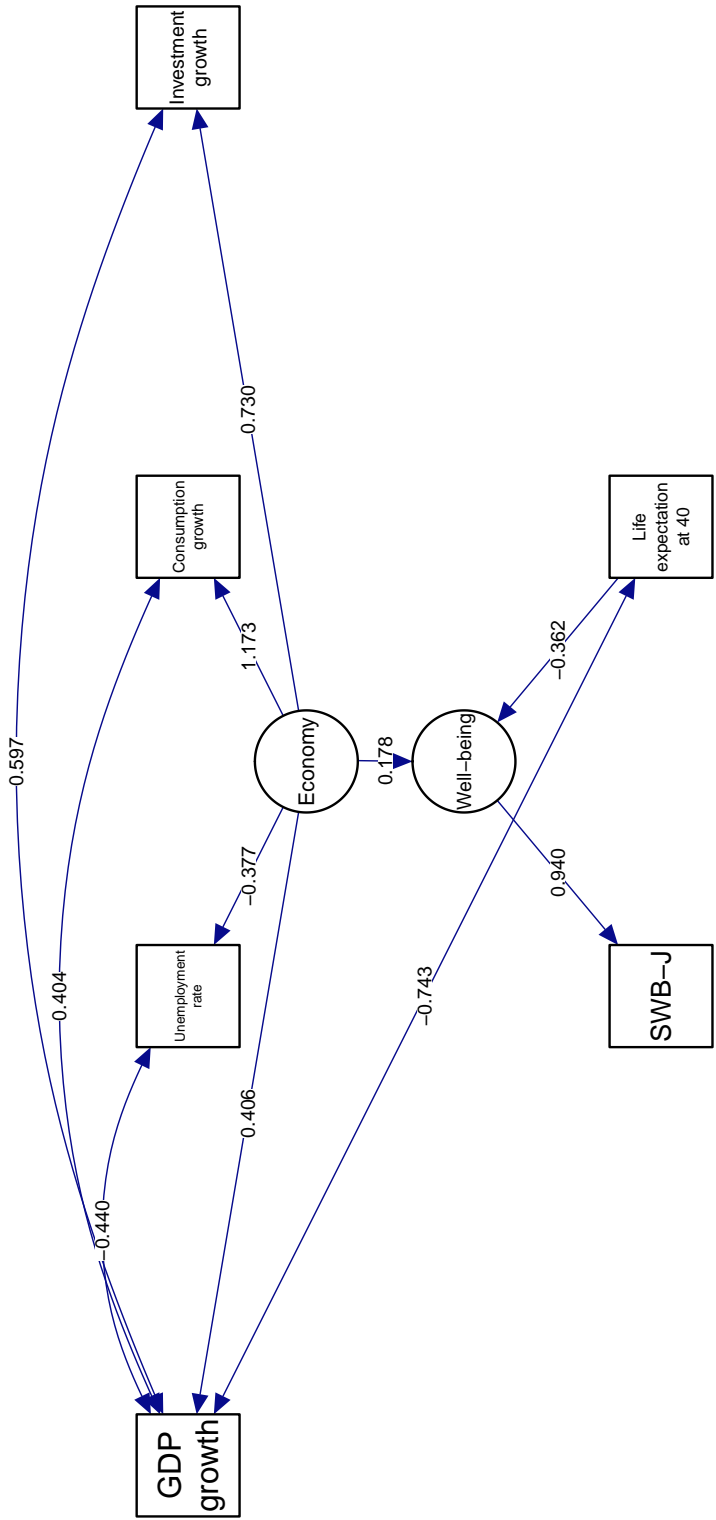

**Figure 3:** Graphical representation of the estimated SEM model for the Japanese data for the period 2015-2018.

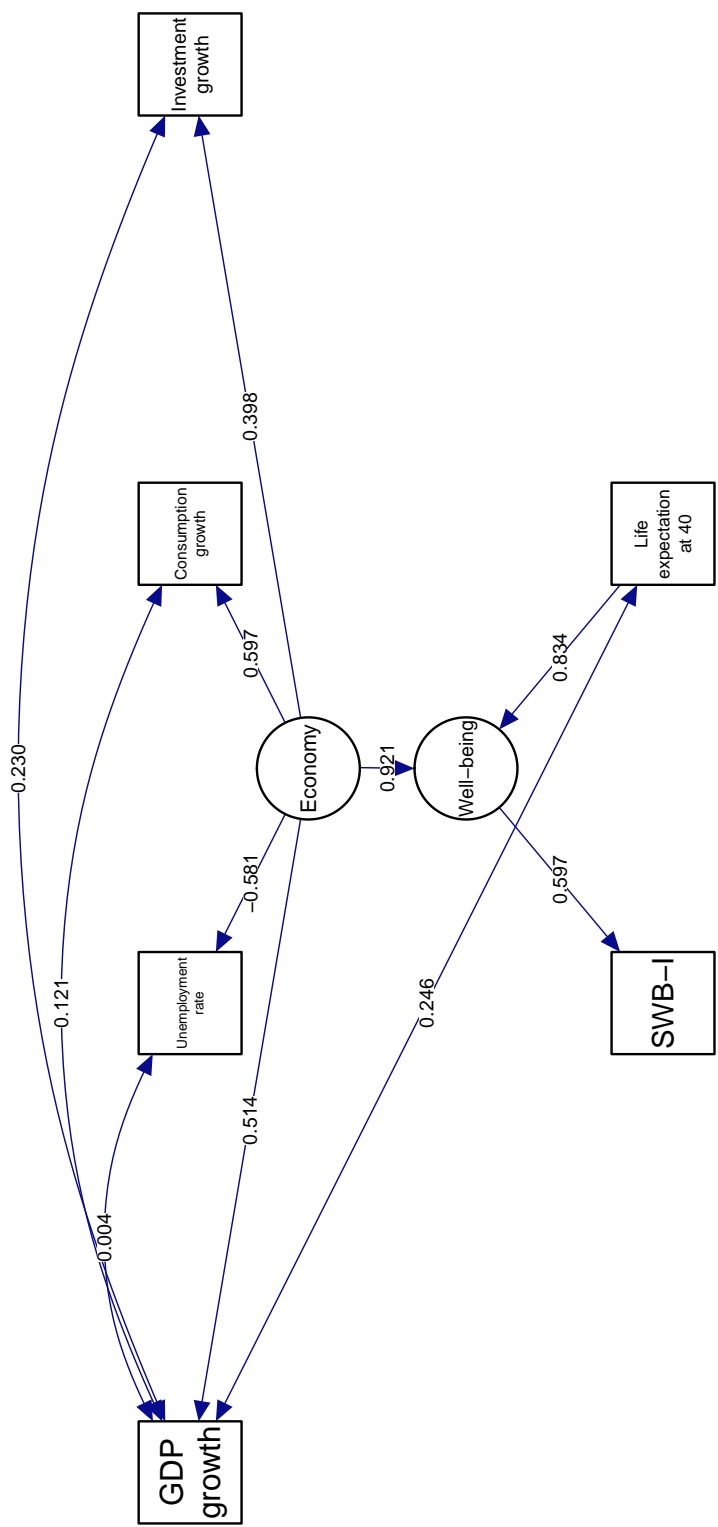

**Figure 4:** Graphical representation of the estimated SEM model for the Italian data for the period 2015-2018.

# 8 Discussion and Limits of the Approach

This was the first attempt to elaborate a subjective well-being index based on Twitter data for Japan. This work shows that the same approach used to construct the analogous subjective well-being indicator for Italy can be transposed to Japan. The limit of the human supervised method is that mother tongue readers are needed to build the training set, but once done it, the iSA algorithm being completely agnostic to language, works seamlessly.

The structural equation model helps emphasizing differences between Japan and Italy, particularly with regard to the relevance of objective elements and circumstances in determining perceived well-being: in fact, it seems that Japanese subjective well-being is less affected by the economic conditions, suggesting that personal, relational and familiar dimensions may play a more important role in the individual and the social well-being sphere.

However, the short time series for Japan may be one of the causes of the non-significance of some coefficients: therefore, future work is foreseen to collect more data in terms of volume and time, in order to test the robustness of the results we obtained. As shows in the appendix, the set of keywords used to select the training set data may be enlarged to encompass more variations on the same concepts. Further, as shown for the Italian case (Iacus et al., 2020b,a, 2019), a local dimension is also important but the collection of data currently available does not allow for this extension of the research. Further data collection with geolocalized data is also needed for future development and application of the method and the indicator.

# Acknowledgments

The data collection has been performed within the Japan Science and Technology Agency CREST (Core Research for Evolutional Science and Technology) project, grant n. JPMJCR14D7. This work was performed also in collaboration with the Waseda Institute of Social Media Data (WISDOM), especially for the part concerning the training set human coding.# Competing and/or conflict of interests

SI is currently on leave from the University of Milan and employed at the Joint Research Centre of the European Commission. SI conducted the statistical analyses while at the University of Milan and wrote the paper while at the Joint Research Centre.

# Author Contributions

SI and GP conceived the project. TC, AH, SI and GP designed the study. TC and AH wrote the intercultural and socio-linguistic analysis. SI and GP did the statistical analyses of the data. TC, AH, SI, and GP interpreted the results and wrote and revised the article.

# References

Abdullah, S., E. L. Murnane, J. M. Costa, and T. Choudhury (2015). Collective smile: Measuring societal happiness from geolocated images. In *Proceedings of the 18th ACM Conference on Computer Supported Cooperative Work & Social Computing*, CSCW '15, pp. 361–374. ACM.

Baker, R., J. M. Brick, N. A. Bates, M. Battaglia, M. P. Couper, J. A. Dever, K. J. Gile, and R. Tourangeau (2013). Summary report of the AAPOR task force on non-probability sampling. *Journal of Survey Statistics and Methodology 1*(2), 90.

Barrington-Leigh, C. and A. Escande (2018). Measuring progress and well-being: A comparative review of indicators. *Social Indicators Research 135*(3), 893˘–925.

Bengio, Y., R. Ducharme, P. Vincent, and C. Janvin (2003). A neural probabilistic language model. *The Journal of Machine Learning Research 3*, 1137–1155.

Bollen, J., B. Gonçalves, G. Ruan, and H. Mao (2011, August). Happiness is assortative in online social networks. *Artif. Life 17*(3), 237–251.

Bollen, J., B. Gonçalves, I. van de Leemput, and G. Ruan (2017). The happiness paradox: your friends are happier than you. *EPJ Data Science 6*(4), 1–10.

Bollen, K. (1989). *tructural Equations with Latent Variables*. New York: Wiley.

Carpi, T. and S. M. Iacus (2020). Is japanese gendered language used on twitter? a large scale study. *Online Journal of Communication and Media Technologies 10*(4), e202024.

Ceron, A., L. Curini, and S. M. Iacus (2016). iSA: A fast, scalable and accurate algorithm for sentiment analysis of social media content. *Information Sciences 367-368*, 105–124.

Curini, L., S. Iacus, and L. Canova (2015). Measuring idiosyncratic happiness through the analysis of Twitter: An application to the Italian case. *Social Indicators Research 121*(2), 525–542.

Deaton, A. (2012). The financial crisis and the wellbeing of americans. *Oxford Economic Papers 64*(1), 1–26.

Deaton, A. and A. A. Stone (2016). Understanding context effects for a measure of life evaluation: how responses matter. *Oxford Economic Papers 68*(4), 861–870.

Diener, E., E. M. Suh, H. Smith, and L. Shao (1995, Jan). National differences in reported subjective well-being: Why do they occur? *Social Indicators Research 34*(1), 7–32.

Dodds, P. S., K. D. Harris, I. M. Kloumann, C. A. Bliss, and C. M. Danforth (2011, 12). Temporal patterns of happiness and information in a global social network: Hedonometrics and twitter. *PLoS ONE 6*(12), 1–26.

Durahim, A. O. and M. Coşkun (2015). #iamhappybecause: Gross national happiness through twitter analysis and big data. *Technological Forecasting and Social Change 99*, 92–105.

Epskamp, S. (2019). *semPlot: Path Diagrams and Visual Analysis of Various SEM Packages' Output*. CRAN. R package version 1.1.2.

Fleurbaey, M. (2009). Beyond gdp: The quest for a measure of social welfare. *Journal of Economic Literature 47*(4), 1029–75.

Ford, B. Q., J. O. Dmitrieva, D. Heller, Y. Chentsova-Dutton, I. Grossmann, M. Tamir, Y. Uchida, B. Koopmann-Holm, V. A. Floerke, M. Uhrig, T. Bokhan, and I. B. Mauss (2015). Culture shapes whether the pursuit of happiness predicts higher or lower well-being. *Journal of Experimental Psychology: General 144*(6), 1053.

Greco, F. and A. Polli (2020, Apr). Security perception and people well-being. *Social Indicators Research TBA*(1), 1–10.

Greyling, T., S. Rossouw, and T. Adhikari (2020). Happiness-lost: Did governments make the right decisions to combat covid-19? https://ideas.repec.org/p/zbw/glodps/556.html.

Hino, A. and R. A. Fahey (2019). Representing the twittersphere: Archiving a representative sample of twitter data under resource constraints. *International Journal of Information Management 48*, 175 – 184.

Iacus, S. M., G. Porro, S. Salini, and E. Siletti (2019). Social networks data and subjective well-being. an innovative measurement for italian provinces. *Scienze Regionali, Italian Journal of Regional Science Speciale*(2019), 667–678.

Iacus, S. M., G. Porro, S. Salini, and E. Siletti (2020a). Controlling for selection bias in social media indicators through official statistics: a proposal. *Journal of Official Statistics 36*(2), 315–338.

Iacus, S. M., G. Porro, S. Salini, and E. Siletti (2020b, Mar). An italian composite subjective well-being index: The voice of twitter users from 2012 to 2017. *Social Indicators Research TBA*(1), 1–19.

Iio, J. (2020). Kawaii in tweets: What emotions does the word describe in social media? In L. Barolli, H. Nishino, T. Enokido, and M. Takizawa (Eds.), *Advances in Networked-based Information Systems*, Cham, pp. 715–721. Springer International Publishing.

Kahneman, D., A. B. Krueger, D. Schkade, N. Schwarz, and A. Stone (2004). Toward national well-being accounts. *American Economic Review 94*(2), 429–434.

Kamijo, K., T. Nasukawa, and H. Kitamura (2016). Personality estimation from japanese text. In *PEOPLES@COLING*.

King, G. (2016). Preface: Big data is not about the data! In R. M. Alvarez (Ed.), *Computational Social Science: Discovery and Prediction*, Chapter 1, pp. 1–10. Cambridge: Cambridge University Press.

Kitayama, S., H. R. Markus, and M. Kurokawa (2000). Culture, emotion, and well-being: Good feelings in japan and the united states. *Cognition and Emotion 14*(1), 93–124. https://doi.org/10.1080/026999300379003.

Kumano, M. (2018). On the concept of well-being in japan: Feeling shiawase as hedonic well-being and feeling ikigai as eudaimonic well-being. *Applied Research in Quality of Life 13*(2), 419–433.

Kuznets, S. (1934). National Income, 1929-1932. In *National Income, 1929-1932*, NBER Chapters, pp. 1–12. National Bureau of Economic Research, Inc.

Landsberger, H. A. (1958). *Hawthorne Revisited: Management and the Worker, its Critics, and Developments in Human Relations in Industry*. Ithaca, N.Y.: Cornell University.

Lim, K. H., K. E. Lee, D. Kendal, E. Rashidi, L.and Naghizade, S. Winter, and M. Vasardani (2018). The grass is greener on the other side: Understanding the effects of green spaces on twitter user sentiments. In *Companion of the The Web Conference 2018*, pp. 275–282. International World Wide Web Conferences Steering Committee.

Luhmann, M. (2017). Using big data to study subjective well-being. *Current Opinion in Behavioral Sciences 18*, 28–33.

Matsumoto, D. (1999). American-japanese cultural differences in judgements of expression intensity and subjective experience. *Cognition and Emotion 13*(2), 201–218.

Mikolov, T., W.-t. Yih, and G. Zweig (2013). Linguistic regularities in continuous space word representations. In *Proceedings of NAACL HLT*.

Miyake, K. (2007). How young japanese express their emotions visually in mobile phone messages: A sociolinguistic analysis. *Japanese Studies 27*(1), 53–72.

Murphy, J., M. W. Link, J. H. Childs, C. L. Tesfaye, E. Dean, M. Stern, J. Pasek, J. Cohen, M. Callegaro, and P. Harwood (2014). Social media in public opinion research. executive summary of the AAPOR task force on emerging technologies in public opinion research. *Public Opinion Quarterly 78*(4), 788–794.

New Economics Foundation (2016). The happy planet index 2016. a global index of sustainable well-being. Technical report, New Economics Foundation.

Novak, P. K., J. Smailović, B. Sluban, and I. Mozetič (2015). Sentiment of emojis. *PLoS ONE 10*, 1–22.

Park, J., Y. M. Baek, and M. Cha (2014). Cross-cultural comparison of nonverbal cues in emoticons on twitter: Evidence from big data analysis. *Journal of Communication 64*(2), 333–354.

Ptaszynski, M., R. Rzepka, K. Araki, and Y. Momouchi (2014). Automatically annotating a five-billion-word corpus of japanese blogs for sentiment and affect analysis. *Computer Speech & Language 28*(1), 38–55.

Quercia, D., J. Ellis, L. Capra, and J. Crowcroft (2012). Tracking gross community happiness from tweets. In *Proceedings of the ACM 2012 Conference on Computer Supported Cooperative Work*, CSCW '12, pp. 965–968. ACM.

Robeyns, I. (2006). The capability approach in practice. *Journal of Political Philosophy 14*(3), 351–376.

Rosseel, Y. (2012). lavaan: An R package for structural equation modeling. *Journal of Statistical Software 48*(2), 1–36.

Rossouw, S. and T. Greyling (2020). Big data and happiness. In K. Zimmermann (Ed.), *Handbook of Labor, Human Resources and Population Economics*, pp. 1–35. Springer.

Rzepka, R., U. Jagla, P. Dybala, and K. Araki (2016). Influence of emoticons and adverbs on affective perception of japanese texts. *Proceedings of the 30th Annual Conference of the Japanese Society for Artificial Intelligence JSAI2016*(1), 3H4OS17b3–3H4OS17b3.

Schwartz, H. A., J. C. Eichstaedt, M. L. Kern, L. Dziurzynski, M. Agrawal, G. J. Park, S. K. Lakshmikanth, S. Jha, M. E. P. Seligman, L. Ungar, and R. E. Lucas (2013). Characterizing geographic variation in well-being using tweets. In *Proceedings of the Seventh International AAAI Conference on Weblogs and Social Media (ICWSM)*.

Schwarz, N. and F. Strack (1999). Reports of subjective well-being: Judgmental processes and their methodological implications. *Well-being: The foundations of hedonic psychology 7*, 61–84.

Scollon, C. N. (2018). Non-traditional measures of subjective well-being and their validity: A review. In E. Diener, S. Oishi, and L. Tay (Eds.), *Handbook of well-being*. Salt Lake City, UT: DEF Publishers.

Sen, A. (1980). Equality of what? In S. M. MacMurrin (Ed.), *The Tanner Lectures on Human Values*, Volume 1, pp. 195–220. Cambridge: Cambridge University Press.

Shoeb, A. and G. de Melo (2020). Are emojis emotional? a study to understand the association between emojis and emotions. *arXiv 1*(1), 1–10.

Stiglitz, J., A. Sen, and J.-P. Fitoussi (2009). Report by the commission on the measurement of economic performance and social progress. Technical report, INSEE.

United Nations Development Programme (2019). Human development report 2019. beyond income, beyond averages, beyond today: Inequalities in human development in the 21st century. Technical report, United Nations Development Programme.

van der Wielen, W. and S. Barrios (2020). Economic sentiment during the covid pandemic: Evidence from search behaviour in the eu. *Journal of Economics and Business TBA*(1), 105970.

Vo, B.-K. H. and N. Collier (2013). Twitter emotion analysis in earthquake situations. *International Journal of Computational Linguistics and Applications 4*, 159–173.

Voukelatou, V., L. Gabrielli, I. Miliou, S. Cresci, R. Sharma, M. Tesconi, and L. Pappalardo (forthcoming). Measuring objective and subjective well-being: dimensions and data sources. *International Journal of Data Science and Analytics TBA*(1), 1–31.

Wong, N., X. Gong, and H. H. Fung (2020). Does valuing happiness enhance subjective well-being? the age-differential effect of interdependence. *Journal of Happiness Studies 21*(1), 1–14.

Yang, C. and P. Srinivasan (2016). Life satisfaction and the pursuit of happiness on twitter. *PLoS ONE 11*(3), 1–30.

Zhao, Y., F. Yu, B. Jing, X. Hu, A. Luo, and K. Peng (2019, Jun). An analysis of well-being determinants at the city level in china using big data. *Social Indicators Research 143*(3), 973–994.

# Appendix: Details on Training Set Construction and Coding Rules

In this appendix we briefly describe the process of coding and we present the list of keywords used to select the tweets for the preparation of the training set. It is important to remark that the selection of the keyword, though not being exhaustive is used across country to keep some relative cultural comparability. Moreover, even though the tweets are selected via the keywords, the actual content is classified by the coder and the keywords are not used in any subsequent set of the analysis. Even the coders do not know from which training set the tweets are coming from as they are asked to code all the eight dimensions (if possible, as explained below) for each post.

For what concerns the application, the mandate for the coders was to consider only self-expressed or individual expression of well-being or own views of the tweeter. In order to obtain an index, the coders should classify the tweets in the categories: positive, neutral, negative and off-Topic. Table 7 is an example of how texts can be distilled into emotions. Table 8 is another possible example.

Unfortunately the natural language, or the real world in general, is more complex than the examples proposed in Tables 7 and 8. For example, the following real tweet (original Japanese version on the left, approximate English translation on the right):

| | |
|---|---|
| 体中が痛い…… | *My body hurts...* |
| 下手くそな証拠だ…… | *Bad shit...* |
| いくちゃんからメール来てたから元気でた | *I was fine because I received an email from Iku-chan.* |
| よし、行動しよ | *OK, act* |

can be categorized under *resilience and self-esteem* (`res`) as *positive*, as well as this one:

| |
|---|
| 意識高い高いと自尊心高い高いしてみたい |

which translates approximately to

| |
|---|
| *"High consciousness and high self-esteem".* |

More complex are tweets like this one:

| |
|---|
| 精神とは控えめに見ても 90 パーセント妄想であって、妄想を自己と切り離す作業を日夜続けることが、初期における人間の精神生活の主なノルマである。 |

which translates approximatively into

| |
|---|
| *The spirit is 90% delusion, even if it is conservative, and continuing the work of separating the delusion from the self, day and night, is the main norm of human mental life in the early days.* |

The above text seems to express a negative view about life which we can arguably classify as *negative* for the component *satisfying life* (`sat`).

Plenty of examples can in fact be produced from real data. And tweets can be classified along one or more dimensions of interest. For example, the one in Figure 5 can be classified as *positive* for the components `emo` and `res` and *negative* for the component `vit`.

| Example (En) | Example (JP) | Classification |
|---|---|---|
| how lucky I am ! | ラッキーだ！ | positive |
| what a beautiful day :) | 美しく晴れ渡った日 | positive |
| finally I passed the exam! | やっと合格した。 | positive |
| there are good and bad people | いい人と悪い人がいる。 | neutral |
| tonight I have a date with my girlfriend <3 | 今晩彼女とデートする予定 <3。 | positive |
| my girlfriend quit me last night | 昨晩彼女に振られちゃった。 | negative |
| I feel sick and I have headache | 風邪を引いて、頭が痛いんだ。 | negative |

**Table 7:** Example of classification rule from fictitious texts with the aim of classifying the *emotional* (`emo`) component of the *personal* well-being.

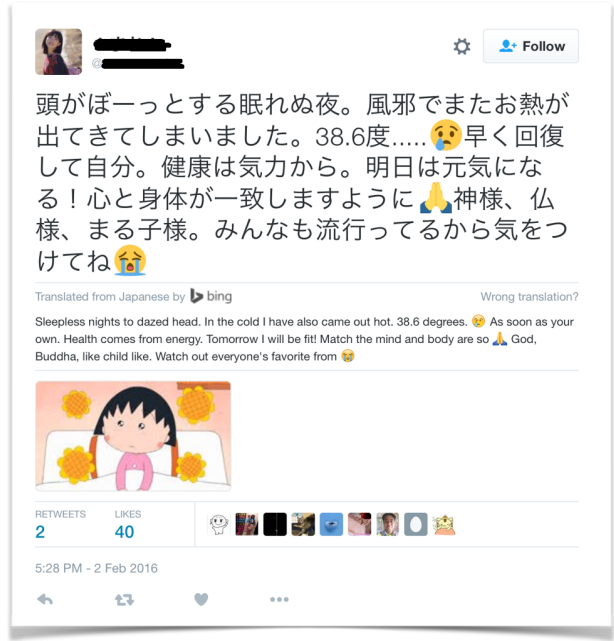

**Figure 5:** Example of real tweet that can be categorized along multiple dimensions: *positive* for the components `emo` and `res` and *negative* for the component `vit`.

| Example (En) | Example (JP) | Classification |
|---|---|---|
| I was very happy that you gave me your support! | 応援してくれてとても嬉しかった！ | positive |
| If only you work sincerely, you will be trusted. | 誠実に働きさえすれば、あなたは信頼されるでしょう。 | neutral |
| It is by no means easy to satisfy everyone. | 全ての者を満足させることをは決して容易ではない。 | neutral |
| I have no great belief in my doctor. | 私は自分の医者をほとんど信頼していません。 | negative |
| She betrayed my trust. | 彼女は自分の信頼を裏切った。 | negative |

**Table 8:** Example of classification rule from fictitious texts with the aim of classifying the *trust and belonging* (`tru`) component of the *social* well-being.

| Italian | English | Japanese |
|---|---|---|
| amore | love | 愛, 好き |
| amicizia | friendship | 友情 |
| emozione | emotion | 感情 |
| sentimento | sentiment | 気持ち, 感じ |
| felice | happy | ハッピー, 喜, 嬉, うれしい |
| felicità | happiness | 幸福, 幸せ |
| lacrime | tears | 涙 |
| gioia | joy | 喜び |
| divertente | funny | おかしい |
| triste | sad | 悲, 不幸 |
| depresso | depressed | 陰気, 愁い, 落ち込む, うつ病 |
| noia | bored | 退屈, うんざり |

**Table 9:** Example of keywords used to select the training set data for the *emotional* (`emo`) component of the *personal* well-being. English is only shown as reference.

| Italian | English | Japanese |
|---|---|---|
| salute | health | 健康　調子　体調　元気 |
| malattia | ill | 病気 |
| famiglia | family | 家族 |
| figli | children | 子ども, 子供 |
| mamma | mother | 母 |
| papà | father | 父 |
| soldi | money | 金 |
| casa | home | 家 |

**Table 10:** Example of keywords used to select the training set data for the *satisfying life* (`sat`) component of the *personal* well-being. English is only shown as reference.

| Italian | English | Japanese |
|---|---|---|
| cinema | cinema | シネマ |
| teatro | theater | 劇場 |
| ristorante | restaurant | レストラン -料亭 |
| palestra | jim | ジム |
| vacanza | holidays | 休日 |
| gita | excursion | 外出 |
| ferie | holidays | 休日 |
| pizza | pizza | pizza -ピザ |
| fitness | fitness | フィットネス |
| jogging | jogging | ジョギング |
| tempo libero | free time | レジャー |
| volontariato | voluntary | 自主的な |
| hobby | hobby | 趣味 |
| club | club | クラブ |
| circolo | social club | 社交クラブ |
| stanco | tired | 疲 |

**Table 11:** Example of keywords used to select the training set data for the *vitality* (`vit`) component of the *personal* well-being. English is only shown as reference.

| Italian | English | Japanese |
|---|---|---|
| fiducia | confidence | 信頼 |
| sicurezza | safety | 安全 |
| paura, timore | fear | 恐怖 |
| capace | capable | 腕利き |
| leader | leader | 棟梁 |
| ottimista | optimistic | 楽観的 |
| ottimismo | optimism | 楽観 |
| futuro | future | 将来 |
| fallimento | failure | 失敗 |
| obiettivo | goal | ターゲット |

**Table 12:** Example of keywords used to select the training set data for the *resilience and self-esteem* (`res`) component of the *personal* well-being. English is only shown as reference.

| Italian | English | Japanese |
|---|---|---|
| libertà | freedom | 自由 |
| autonomia | autonomy | 自治 |
| significato | meaning | 意味 |
| imparare | learn | 学ぶ |

**Table 13:** Example of keywords used to select the training set data for the *positive functioning* (`fun`) component of the *personal* well-being. English is only shown as reference.

| Italian | English | Japanese |
|---|---|---|
| aiuto | help | 助けて |
| vicini casa | neighbors | 隣人 -助 |
| rispetto | respect | 尊敬 |

**Table 14:** Example of keywords used to select the training set data for the *trust and belonging* (`tru`) component of the *social* well-being. English is only shown as reference.

| Italian | English | Japanese |
|---|---|---|
| famiglia | family | 家族 |
| figli | children | 子供 |
| mamma | mother | 母 |
| papà | father | 父 |
| fratello | brother | 兄 |
| sorella | sister | 妹, 姉 |
| amici | friends | 友人, 友達 |
| marito | husband | 夫 |
| moglie | wife | 妻 |
| parenti | relatives | 親族, 身内 |
| solitudine | loneliness | 孤独 |

**Table 15:** Example of keywords used to select the training set data for the *relationship* (`rel`) component of the *social* well-being. English is only shown as reference.

| Italian | English | Japanese |
|---|---|---|
| lavoro | job, work | 作業, 職, 仕事, 職業, 作業 |
| carriera | career | キャリア, 経歴, 来歴, 閲歴 |
| collega | colleague | 同僚 |
| ufficio | office | 事務所 |
| tempo lavoro | working time | 労働時間 |
| stress | stress | ストレス |
| disoccupato | unemployed | 失業者 |
| disoccupazione | unemployment | 失業 |
| contratto lavoro | job contract | 雇用契約 |
| stipendio | salary | 給与 |
| merito | merit | 実力, 有功 |

**Table 16:** Example of keywords used to select the training set data for the *quality of job* (`wor`) component of the well-being *at work*. English is only shown as reference.